\documentstyle[12pt,a4,here, epic,eepic]{article}

\makeatletter
   \def\@cite#1#2{\leavevmode\hbox{$^{\mbox{\the\scriptfont0 #1}}$}}
\makeatother

     \def\sD{\scriptscriptstyle D}

\makeatletter
\@addtoreset{equation}{subsection}

\makeatother

\makeatletter
\long\def\@makecaption#1#2{%
   \vskip 10\p@
   \setbox\@tempboxa\hbox{#1\ \ #2}%
   \ifdim \wd\@tempboxa >\hsize
   #1\ \ #2\par        %
      \else
   \hbox to\hsize{\hfil\box\@tempboxa\hfil}%
   \fi}
\def\fnum@figure{Fig. \thefigure}
\makeatother

\begin{document}

\begin{flushright}
\begin{tabular}{l}
HUPD-9512 \hspace{1em}\\
May 1995
\end{tabular}
\end{flushright}

\vspace{1.5cm}

\begin{center}
{\Large \bf
A Soluble Model of Four-Fermion Interactions\\[5mm]
in de Sitter Space} \\[2cm]
\normalsize
T.~Inagaki
\footnote{e-mail : inagaki@theo.phys.sci.hiroshima-u.ac.jp},
S.~Mukaigawa and T.~Muta
\footnote{Supported in part by Monbusho Grant-in-Aid for Scientific
Research (C) under contract \\No.04640301.
}\\[1cm]
{\it
Department of Physics, Hiroshima University, \\
Higashi-Hiroshima, Hiroshima 739, Japan \\[3cm]
}
\end{center}

\begin{abstract}
\vglue 0.7cm
We consider the theory of four-fermion interactions
with N-component fermions in de Sitter space.
It is found that the effective potential for a composite operator
in the theory is calculable in the leading order of the 1/N
expansion.
The resulting effective potential is analyzed by varying both the
four-fermion coupling constant and the curvature of the space-time.
The critical curvature at which the dynamically generated
fermion mass disappears is found to exist and is calculated analytically.
The dynamical fermion mass is expressed as a function of the space-time
curvature.
\end{abstract}

\newpage

%\pacs{11.15.Pg,11.30.Qc}

\renewcommand{\thesubsection}{\arabic{subsection}}
\renewcommand{\thesubsubsection}
   {\arabic{subsection}.\arabic{subsubsection}}
\baselineskip=24pt

According to the well-accepted scenario of early universe it is believed
that the GUT phase is broken down to the QCD and electroweak phase through
the symmetry breaking due to the Higgs mechanism. At this era the quantum
gravity plays a minor role while the space-time curvature due to the external
strong gravity is important in triggering the phase transition. Thus it is
of interest to consider quantum field theory in curved space-time in
connection with physics in the very early universe.\cite{AL}
On the other hand the
Higgs mechanism is often explained as a dynamical effect due to the emergence
of the composite Higgs field. A typical example of such a model is the
technicolor model.\cite{TC}
In this regard it is interesting to deal with a quantum
field theory with the composite Higgs field. The four-fermion interaction
theory is one of the prototype models of the composite Higgs theory.\cite{NJL}

Under these circumstances we find it useful to discuss the phase
structure of the four-fermion theory in curved space-time. Since the phase
transition is a nonperturbative phenomenon, we need to use the method free
from the perturbative approach and also to avoid any approximation in dealing
with the space-time curvature, e. g. a weak curvature approximation in which
we rely on an expansion in powers of the curvature.

In the present paper we adopt the 1/N expansion method as a
nonperturvative approach and try to find the effective potential without
making any approximation in the space-time curvature. For this purpose we
restrict ourselves to the specific space-time, i. e. the de Sitter space-time,
and calculate the effective potential for the composite operator made
of a fermion-antifermion pair. By the use of the
effective potential we shall argue the symmetry breaking caused by the
curvature effect.

We consider the theory in the curved space-time defined by the action,
\begin{equation}
     S=\int d^{\sD}x\sqrt{-g(x)}
     \left[
          \sum^{N}_{k=1}\bar{\psi}_{k}\gamma^{\mu}\nabla_{\mu}\psi_{k}
          +\frac{\lambda_{0}}{2N}
          \left(\sum^{N}_{k=1}\bar{\psi}_{k}\psi_{k}\right)^{2}
     \right]\, ,
\label{act1}
\end{equation}
where index $k$ represents the flavor of the fermion field $\psi$,
$N$ is the number of fermion species,
$g$ the determinant of the space-time metric $g_{\mu\nu}$,
$\gamma^{\mu}$ the Dirac matrix in the curved space and
$\nabla_{\mu}\psi$ the covariant derivative of the fermion field $\psi$.
Throughout the paper we work in arbitrary space-time dimension $D$.
For simplicity we neglect the flavor index and the
summation on it.
Our notation is
the $(+,+,+)$ convention as defined
in the book by Misner, Thorne and Wheeler.\cite{MTW}
In the following calculations it is more convenient to introduce
auxiliary field $\sigma$ and to consider the action,
\begin{equation}
     S^{\prime}=\int d^{\sD}x\sqrt{-g(x)}
     \left[
          \bar{\psi}\gamma^{\mu}\nabla_{\mu}\psi
          +\bar{\psi}\sigma\psi
          -\frac{N}{2\lambda_{0}}\sigma^{2}
     \right]\, .
\label{act2}
\end{equation}
It is well-known that the physics described by action ($\ref{act2}$)
is equivalent to that described by action ($\ref{act1}$).\cite{GN}
We are interested in estimating the effective potential for the composite
field $\bar{\psi}\psi$ which is essentially the same as the auxiliary field
$\sigma$.
We calculate the effective potential $V(\sigma)$
for field $\sigma$ in the theory defined by action $S^{\prime}$
in Eq.(\ref{act2}).

We start with the generating functional given by
\begin{equation}
     Z=\int [d\psi d\bar{\psi}d\sigma] \exp(iS)\, .
\label{gne1}
\end{equation}
Performing the integration over the fermion fields $\psi$ and $\bar{\psi}$
we obtain
\begin{equation}
     Z=\int [d\sigma] \exp(i N S_{eff})\, ,
\label{gne2}
\end{equation}
where $S_{eff}$ is given by
\begin{equation}
     S_{eff}=-\int d^{\sD}x\sqrt{-g(x)}
           \frac{1}{2\lambda_{0}}\sigma^{2}
          -i\mbox{Tr ln}(\gamma^{\mu}\nabla_{\mu}+\sigma)\, .
\end{equation}
In the following argument we restrict ourselves to the
space-time where the path integrals (\ref{gne1}) and (\ref{gne2})
are well-defined.
As it may be seen that
quantum corrections relevant to the auxiliary field $\sigma$ are
only of higher order in the $1/N$ expansion,
we realize that the field $\sigma$ is regarded as a classical field
in the leading order of the $1/N$ expansion and the
effective potential $V(\sigma)$ reads
\begin{equation}
     V(\sigma)=\frac{1}{2\lambda_{0}}\sigma^{2}
     +i\mbox{tr}\langle x \mid
      \ln \frac{\gamma^{\mu}\nabla_{\mu}+\sigma}{\gamma^{\mu}\nabla_{\mu}}
      \mid x \rangle +\mbox{O}(1/N)\, ,
\label{v:eff}
\end{equation}
where the potential is normalized so that $V(0)=0$ and
variable $\sigma$ is independent of the space-time
coordinate.

To estimate the second term on the right-hand side of Eq.(\ref{v:eff})
we adopt the Schwinger proper time method, i.e.
\begin{equation}
     V(\sigma)= \frac{1}{2\lambda_{0}}\sigma^{2}
     +i\mbox{tr}\int^{\sigma}_{0}ds\ G(x,x;s)\, ,
\label{v:desitter}
\end{equation}
where $G(x,y;s)$ is defined by
\begin{equation}
     G(x,y;s)=\langle x |(\gamma^{\mu}\nabla_{\mu}+s)^{-1}| y \rangle\, .
\end{equation}
The function $G(x,y;s)$ satisfies the differential equation
\begin{equation}
     (\gamma^{\mu}\nabla_{\mu}+s)G(x,y;s)=\delta^{\sD}(x,y)\, ,
\label{eq:gree}
\end{equation}
where $\delta^{\sD}(x,y)$ is the Dirac delta function in the curved
space.
With this equation we recognize that $G(x,y,;s)$ is essentially equal to
the Green function for the massive free fermion with mass $s$ in the
curved space-time.

Our problem of calculating the effective potential for the composite
field in the leading order of the $1/N$ expansion for the
four-fermion theory in the curved space-time is now reduced to the
problem of finding the Green function for the free massive fermion
in the curved space-time.
Fortunately it is known that the Green function for the free massive fermion
is exactly calculable in de Sitter space.\cite{CW,ESM,ALL}
We then restrict ourselves to de Sitter space and closely follow the
method developed by Candelas and Raine.\cite{ESM}
The de Sitter space is a maximally symmetric curved space-time
with constant curvature.
The de Sitter space of $D$ dimensions is represented as a hyperboloid
\begin{equation}
     r^2=-{\xi_{0}}^{2}+{\xi_{1}}^{2}+\cdots+{\xi_{\sD}}^{2}
\label{def:desitter}
\end{equation}
embedded in the $D+1$-dimensional Minkowski space.
The metric in de Sitter space with variable $\xi_{\mu}$
$(\mu =0,1,2,\cdots ,D-1)$ is given by
\begin{equation}
     g_{\mu\nu}=\eta_{\mu\nu}+\frac{\xi_{\mu}\xi_{\nu}}{r^{2}-\xi^{2}}\, .
\end{equation}
The space-time curvature $R$ for de Sitter space reads
\begin{equation}
     R=D(D-1) r^{-2}\, .
\label{corr:ra}
\end{equation}
We concentrate ourselves on solving Eq.(\ref{eq:gree})
in de Sitter space with the Dirac delta function,
\begin{equation}
     \delta(x,x^{\prime}) =
     \frac{ie^{-i\sD\pi/4}}{(4\pi\epsilon)^{\sD/2}}
     e^{i\frac{\sigma}{2\epsilon}} \sqrt{-g} \, ,
\end{equation}
where $\epsilon$ is a parameter which is set equal to zero after calculations.
We first note that the Green function $G(x,y;s)$
in de Sitter space depends on two variables $x$ and $y$
through
\begin{equation}
     \sigma(x,x^{\prime})
     = \frac{1}{2}(\vec{\xi}-\vec{\xi}^{\prime})^{2}  \, ,
\end{equation}
according to the maximal symmetry of de Sitter space.
As is pointed out by Candelas and Raine,\cite{ESM} the Green function
$G(x,y;s)$ may be decomposed into invariant amplitudes
$A(\sigma)$ and $B(\sigma)$ in the following way:
\begin{equation}
     G(x,y ;s) = H(x,y)\Phi(x,y)  \, ,
     \label{eqn:bunkai}
\end{equation}
with $\Phi(x,x) = unit\,\, matrix \, ,$
and
\begin{equation}
     H(x,y) = A(\sigma) + B(\sigma)\sigma_{;\alpha}\gamma^{\alpha}\, .
\label{eqn:hh}
\end{equation}
We substitute the expression (\ref{eqn:bunkai}) with Eq.(\ref{eqn:hh})
into Eq.(\ref{eq:gree}) and take
the trace on both side of the equation.
After some algebra we obtain for the invariant
amplitude $A(\sigma)$
\begin{eqnarray}
     &&z(z-1)\frac{d^{2}A}{dz^{2}}+D(z-\frac{1}{2})\frac{dA}{dz}
     +r^{2}(\frac{i}{2r}(2-D)+s)(\frac{i}{2r}D+s)A
                                            \nonumber \\
     &=&r^{2}(\frac{i}{2r}(2-D)+s)
     ie^{-i\sD\pi/4} \frac{e^{i\sigma/2\epsilon}}{(4\pi\epsilon)^{\sD/2}} \, ,
\label{eq:A}
\end{eqnarray}
where we made the change of variable:
$\displaystyle z=\frac{\sigma}{2r^{2}} \, .$
Eq.(\ref{eq:A}) is the hypergeometric differential equation
whose solution is given by
\begin{equation}
     A = \frac{a r^{1-\sD}}{(4\pi)^{\sD/2}}
     \frac{\Gamma(a)\Gamma(b)}{\Gamma(D/2)}
     F(a,b,D/2;1-\frac{\sigma}{2r^{2}})   \, ,
\label{sol:a}
\end{equation}
where $F(a,b,c;z)$ is the hypergeometric function \cite{ZETA} of
variable $z$ with parameters $a,b,c$ and
\begin{equation}
\left\{
\begin{array}{ll}
     \displaystyle a = \frac{D-2}{2}+isr \, , \\[4mm]
     \displaystyle b = \frac{D}{2}-isr \, .
\end{array}
\right.
\end{equation}
{}From the solution (\ref{sol:a}) $\mbox{tr} G(x,x;s)$ may be easily
obtained with recourse to Eq.(\ref{eqn:bunkai}) with
Eq.(\ref{eqn:hh}):
\begin{equation}
     \mbox{tr}G(x,x;s) = \frac{a r^{1-\sD}}{(4\pi)^{\sD/2}}
             \frac{\Gamma(a)\Gamma(b)\Gamma(-a-b+D/2)}
            {\Gamma(-a+D/2)\Gamma(-b+D/2)} \mbox{tr\boldmath$1$}  \, .
\label{g:desitter}
\end{equation}
Here by tr{\boldmath$1$} we mean the trace of the unit Dirac matrix.
Inserting Eq.(\ref{g:desitter}) into Eq.(\ref{v:desitter}) our final
expression of the effective potential is obtained.
\begin{equation}
     V(\sigma)=\frac{1}{2\lambda_{0}}\sigma^{2}-\int^{\sigma}_{0}ds
     \frac{s r^{2-\sD}}{(4\pi)^{\sD/2}}
     \frac{\displaystyle \Gamma\left(\frac{D}{2}+i s r\right)
     \Gamma\left(\frac{D}{2}-i s r\right)}
     {\displaystyle \Gamma\left(1+i s r\right)
     \Gamma\left(1-i s r\right)}
     \Gamma\left(1-\frac{D}{2}\right) \mbox{tr\boldmath$1$}  \, .
\label{v:nonren}
\end{equation}
Equation (\ref{v:nonren}) is an exact expression of the
effective potential for the model of four-fermion interactions
in de Sitter space in the leading order of the 1/N expansion.

The effective potential (\ref{v:nonren}) is divergent in two
and four dimensions.
In four dimensions the theory is nonrenormalizable and so we
restrict ourselves to the case $D<4$.
In two dimensions the theory
is renormalizable so that the effective potential is given a finite
expression by the renormalization procedure.
All the divergent terms in the effective potential are essentially
of the same form as those in the flat space-time.
(Note that according to our normalization $V(0)=0$ divergences
associated with the vacuum energy are absent.)
Thus the necessary renormalization can be performed in the vanishing
curvature limit, i.e. in the Minkowski space.
Taking the limit $R\rightarrow 0$ in Eq.(\ref{v:nonren})
we obtain the effective potential $V_{0}(\sigma)$
in the $D$-dimensional Minkowski space: \cite{IKM}
\begin{equation}
     V_{0}(\sigma)=\frac{1}{2\lambda_{0}}\sigma^{2}
     -\frac{1}{(4\pi)^{\sD /2}D}\Gamma\left(
     1-\frac{D}{2}\right)\sigma^{\sD} \mbox{tr\boldmath$1$}  \, .
\label{v0}
\end{equation}
Here the following formula has been utilized,\cite{ZETA}
\begin{equation}
     \frac{\Gamma(\frac{D}{2}+isr)
      \Gamma(\frac{D}{2}-isr)}
     {\Gamma(1 + isr)
      \Gamma(1 - isr)}
     =
     \frac{|\Gamma(\frac{D}{2}+isr)|^{2}}
     {|\Gamma(1 + isr)|^{2}}
     \sim
     ({|s|r})^{\sD-2}
     \mbox{\hspace{3ex}}(r\rightarrow \infty)\, .
\end{equation}
We impose the renormalization condition
\begin{equation}
     \left.
     \frac{\partial^{2}V_{0}(\sigma)}{\partial \sigma^{2}}
     \right|_{\sigma = \sigma_{0}}
     =\frac{\sigma_{0}^{\sD-2}}{\lambda_{r}}\, ,
\label{cond:ren}
\end{equation}
with $\sigma_{0}$ the renormalization scale and $\lambda_{r}$
the renormalized dimensionless coupling constant.
{}From Eq.(\ref{cond:ren}) we find
\begin{equation}
     \frac{1}{\lambda_{0}}=\frac{\sigma_{0}^{\sD-2}}{\lambda_{r}}
     +\frac{D-1}{(4\pi)^{\sD /2}}\Gamma\left(
     1-\frac{D}{2}\right)\sigma_{0}^{\sD-2} \mbox{tr\boldmath$1$} \, .
\label{coup:ren}
\end{equation}
Replacing $\lambda_{0}$ by $\lambda_{r}$ in Eq.(\ref{v0})
we obtain the renormalized effective potential in the
Minkowski space,
\begin{eqnarray}
     \frac{V_{0}(\sigma)}{\sigma_{0}^{\sD}} =
     \frac{1}{2\lambda_{r}}\left(\frac{\sigma}{\sigma_{0}}\right)^{2}
     &+&\frac{D-1}{2(4\pi)^{\sD/2}} \Gamma\left(1-\frac{D}{2}\right)
     (\frac{\sigma}{\sigma_{0}})^{2}
      \mbox{tr\boldmath$1$}  \nonumber \\
     &-&\frac{1}{(4\pi)^{\sD/2} D} \Gamma\left(1-\frac{D}{2}\right)
     \left(\frac{\sigma}{\sigma_{0}}\right)^{\sD}
      \mbox{tr\boldmath$1$}  \, .
\end{eqnarray}
Through the gap equation
    $ \displaystyle\left.\frac{\partial V_{0}(\sigma)}
     {\partial \sigma}\right|_{\sigma=m_{0}} = 0$
we obtain the dynamical fermion mass
\begin{equation}
     m_{0}
     =\sigma_{0}\left[ \frac{(4\pi)^{\sD/2}}
     {\displaystyle\Gamma\left(1-\frac{D}{2}\right) \mbox{tr\boldmath$1$}}
     \left(\frac{1}{\lambda_{r}}
     - \frac{1}{\lambda_{cr}}\right)\right]^{1/(D-2)}
     \, ,
\end{equation}
if $\lambda_{r} > \lambda_{cr}$ where
\begin{equation}
     \frac{1}{\lambda_{cr}} \equiv \frac{1-D}{(4\pi)^{\sD/2}}
     \Gamma\left(1-\frac{D}{2}\right) \mbox{tr\boldmath$1$}  \, .
\label{l:cr}
\end{equation}
Substituting the renormalized coupling constant $\lambda_{r}$
into the Eq.(\ref{v:nonren}) we find the renormalized expression of the
effective potential $V(\sigma)$ in de Sitter space,
\begin{eqnarray}
     V(\sigma)&=&\left[\frac{1}{2\lambda_{r}}
     +\frac{D-1}{2(4\pi)^{\sD /2}}\Gamma\left(
     1-\frac{D}{2}\right)
      \mbox{tr\boldmath$1$}\right]\sigma_{0}^{\sD-2}\sigma^{2} \nonumber \\
     &&-\int^{\sigma}_{0}ds
     \frac{s r^{2-\sD}}{(4\pi)^{\sD/2}}
     \frac{\displaystyle \Gamma\left(\frac{D}{2}+i s r\right)
     \Gamma\left(\frac{D}{2}-i s r\right)}
     {\displaystyle \Gamma\left(1+i s r\right)
     \Gamma\left(1-i s r\right)}
     \Gamma\left(1-\frac{D}{2}\right) \mbox{tr\boldmath$1$}  \, .
\label{v:ren}
\end{eqnarray}
Note that Eq.(\ref{v:ren}) reduces to
\begin{eqnarray}
\frac{V^{D=2}(\sigma)}{\sigma_{0}^{2}} = \left[\frac{1}{2\lambda_{r}}\right.
&+& \left.\frac{1}{2\pi}\left(\ln\frac{1}{r\sigma_{0}}-1\right) \right]
    \left(\frac{\sigma}{\sigma_{0}}\right)^{2} \nonumber \\
&+& \frac{1}{{\sigma_{0}}^{2}} \int^{\sigma}_{0} ds s
     \left[ \psi(1 + isr)
       +\psi(1 - isr) \right]\, ,
\end{eqnarray}
in two dimensions which is different from the expression obtained
in Ref.11 and is consistent with the one in Ref.12.
In Fig.1 the behavior of the effective potential given by Eq.(\ref{v:ren})
is presented in the case of $D=2.5$ for several typical values
of the curvature.
It is observed in Fig.\ref{fig:pot} that, if $\lambda_{r} < \lambda_{cr}$,
the theory is always in the symmetric phase as the curvature changes
while, if   $\lambda_{r} > \lambda_{cr}$, the symmetry restoration
takes place as the curvature exceeds its critical value
$1/r_{cr}=\sigma_{0}/2.74$.
This observation remains true if the space-time dimension is arbitrarily
changed.
\begin{figure}
% GNUPLOT: LaTeX picture using EEPIC macros
\setlength{\unitlength}{0.240900pt}
\begin{picture}(1500,900)(0,0)
\tenrm
\thicklines \path(220,113)(240,113)
\thicklines \path(1436,113)(1416,113)
\put(198,113){\makebox(0,0)[r]{0}}
\thicklines \path(220,189)(240,189)
\thicklines \path(1436,189)(1416,189)
\put(198,189){\makebox(0,0)[r]{0.0005}}
\thicklines \path(220,266)(240,266)
\thicklines \path(1436,266)(1416,266)
\put(198,266){\makebox(0,0)[r]{0.001}}
\thicklines \path(220,342)(240,342)
\thicklines \path(1436,342)(1416,342)
\put(198,342){\makebox(0,0)[r]{0.0015}}
\thicklines \path(220,419)(240,419)
\thicklines \path(1436,419)(1416,419)
\put(198,419){\makebox(0,0)[r]{0.002}}
\thicklines \path(220,495)(240,495)
\thicklines \path(1436,495)(1416,495)
\put(198,495){\makebox(0,0)[r]{0.0025}}
\thicklines \path(220,571)(240,571)
\thicklines \path(1436,571)(1416,571)
\put(198,571){\makebox(0,0)[r]{0.003}}
\thicklines \path(220,648)(240,648)
\thicklines \path(1436,648)(1416,648)
\put(198,648){\makebox(0,0)[r]{0.0035}}
\thicklines \path(220,724)(240,724)
\thicklines \path(1436,724)(1416,724)
\put(198,724){\makebox(0,0)[r]{0.004}}
\thicklines \path(220,801)(240,801)
\thicklines \path(1436,801)(1416,801)
\put(198,801){\makebox(0,0)[r]{0.0045}}
\thicklines \path(220,877)(240,877)
\thicklines \path(1436,877)(1416,877)
\put(198,877){\makebox(0,0)[r]{0.005}}
\thicklines \path(220,113)(220,133)
\thicklines \path(220,877)(220,857)
\put(220,68){\makebox(0,0){0}}
\thicklines \path(524,113)(524,133)
\thicklines \path(524,877)(524,857)
\put(524,68){\makebox(0,0){0.05}}
\thicklines \path(828,113)(828,133)
\thicklines \path(828,877)(828,857)
\put(828,68){\makebox(0,0){0.1}}
\thicklines \path(1132,113)(1132,133)
\thicklines \path(1132,877)(1132,857)
\put(1132,68){\makebox(0,0){0.15}}
\thicklines \path(1436,113)(1436,133)
\thicklines \path(1436,877)(1436,857)
\put(1436,68){\makebox(0,0){0.2}}
\thicklines \path(220,113)(1436,113)(1436,877)(220,877)(220,113)
\put(45,945){\makebox(0,0)[l]{\shortstack{$V/\sigma_{0}^{2.5}$}}}
\put(828,-22){\makebox(0,0){$\sigma/\sigma_{0}$}}
\put(828,220){\makebox(0,0)[l]{$1/r=0$}}
\put(828,633){\makebox(0,0)[l]{$1/r=2/r_{cr}$}}
\thinlines \path(220,113)(220,113)(221,113)(222,113)(223,113)(225,113)
(227,113)(229,113)(232,113)(234,113)(239,113)(243,113)(248,113)(257,114)
(267,114)(276,115)(294,116)(313,118)(332,121)(369,126)(406,134)(443,143)
(481,154)(518,167)(555,181)(592,197)(629,215)(667,235)(704,256)(741,279)
(778,304)(816,331)(853,360)(890,390)(927,423)(964,457)(1002,494)(1039,532)
(1076,572)(1113,615)(1151,659)(1188,706)(1225,755)(1262,805)(1300,858)
(1312,877)
\thinlines \path(220,113)(220,113)(221,113)(222,113)(223,113)(225,113)
(227,113)(229,113)(232,113)(234,113)(239,113)(243,113)(248,113)(257,114)
(267,114)(276,114)(294,115)(313,116)(332,118)(369,122)(406,127)(443,133)
(481,140)(518,149)(555,159)(592,171)(629,184)(667,199)(704,215)(741,233)
(778,252)(816,274)(853,297)(890,322)(927,348)(964,377)(1002,408)(1039,440)
(1076,475)(1113,512)(1151,550)(1188,591)(1225,634)(1262,679)(1300,727)
(1337,776)(1374,828)(1407,877)
\thinlines \path(220,113)(220,113)(221,113)(222,113)(223,113)(225,113)
(227,113)(229,113)(232,113)(234,113)(239,113)(243,113)(248,113)(257,114)
(267,114)(276,114)(294,116)(313,117)(332,119)(369,124)(406,130)(443,137)
(481,146)(518,156)(555,168)(592,181)(629,195)(667,212)(704,229)(741,249)
(778,270)(816,293)(853,317)(890,344)(927,372)(964,402)(1002,434)(1039,468)
(1076,504)(1113,542)(1151,582)(1188,624)(1225,668)(1262,715)(1300,763)
(1337,814)(1374,867)(1380,877)
\thinlines \path(220,113)(220,113)(221,113)(222,113)(223,113)(225,113)
(227,113)(228,113)(229,113)(234,113)(236,113)(239,113)(248,113)(253,113)
(257,113)(276,114)(285,114)(294,114)(332,116)(350,118)(369,120)(406,124)
(443,129)(481,136)(518,144)(555,154)(592,164)(629,177)(667,191)(704,207)
(741,224)(778,243)(816,264)(853,286)(890,311)(927,337)(964,365)(1002,395)
(1039,427)(1076,462)(1113,498)(1151,536)(1188,576)(1225,619)(1262,664)
(1300,711)(1337,760)(1374,812)(1411,866)(1419,877)
\end{picture}
\vglue 2ex
\hspace*{4em}(a)$\lambda_{r}=0.9\lambda_{cr}$
\vglue 6ex
% GNUPLOT: LaTeX picture using EEPIC macros
\setlength{\unitlength}{0.240900pt}
\begin{picture}(1500,900)(0,0)
\tenrm
\thicklines \path(220,113)(240,113)
\thicklines \path(1436,113)(1416,113)
\put(198,113){\makebox(0,0)[r]{-0.0002}}
\thicklines \path(220,209)(240,209)
\thicklines \path(1436,209)(1416,209)
\put(198,209){\makebox(0,0)[r]{-0.0001}}
\thicklines \path(220,304)(240,304)
\thicklines \path(1436,304)(1416,304)
\put(198,304){\makebox(0,0)[r]{0}}
\thicklines \path(220,400)(240,400)
\thicklines \path(1436,400)(1416,400)
\put(198,400){\makebox(0,0)[r]{0.0001}}
\thicklines \path(220,495)(240,495)
\thicklines \path(1436,495)(1416,495)
\put(198,495){\makebox(0,0)[r]{0.0002}}
\thicklines \path(220,591)(240,591)
\thicklines \path(1436,591)(1416,591)
\put(198,591){\makebox(0,0)[r]{0.0003}}
\thicklines \path(220,686)(240,686)
\thicklines \path(1436,686)(1416,686)
\put(198,686){\makebox(0,0)[r]{0.0004}}
\thicklines \path(220,782)(240,782)
\thicklines \path(1436,782)(1416,782)
\put(198,782){\makebox(0,0)[r]{0.0005}}
\thicklines \path(220,877)(240,877)
\thicklines \path(1436,877)(1416,877)
\put(198,877){\makebox(0,0)[r]{0.0006}}
\thicklines \path(220,113)(220,133)
\thicklines \path(220,877)(220,857)
\put(220,68){\makebox(0,0){0}}
\thicklines \path(524,113)(524,133)
\thicklines \path(524,877)(524,857)
\put(524,68){\makebox(0,0){0.05}}
\thicklines \path(828,113)(828,133)
\thicklines \path(828,877)(828,857)
\put(828,68){\makebox(0,0){0.1}}
\thicklines \path(1132,113)(1132,133)
\thicklines \path(1132,877)(1132,857)
\put(1132,68){\makebox(0,0){0.15}}
\thicklines \path(1436,113)(1436,133)
\thicklines \path(1436,877)(1436,857)
\put(1436,68){\makebox(0,0){0.2}}
\thicklines \path(220,113)(1436,113)(1436,877)(220,877)(220,113)
\put(45,945){\makebox(0,0)[l]{\shortstack{$V/\sigma_{0}^{2.5}$}}}
\put(828,-22){\makebox(0,0){$\sigma/\sigma_{0}$}}
\put(1041,228){\makebox(0,0)[l]{$1/r=0$}}
\put(463,591){\makebox(0,0)[l]{$1/r=2/r_{cr}$}}
\thinlines \path(220,304)(220,304)(221,304)(222,304)(223,304)(225,304)
(227,304)(229,304)(232,304)(234,304)(239,304)(243,304)(248,305)(257,305)
(267,306)(276,306)(294,308)(313,311)(332,314)(369,322)(406,332)(443,344)
(481,359)(518,376)(555,396)(592,418)(629,443)(667,471)(704,501)(741,535)
(778,572)(816,612)(853,656)(890,703)(927,754)(964,809)(1002,868)(1007,877)
\thinlines \path(220,304)(220,304)(221,304)(222,304)(223,304)(225,304)
(227,304)(229,304)(232,304)(234,304)(239,304)(243,304)(248,304)(257,303)
(276,302)(294,301)(332,297)(369,292)(406,287)(443,280)(481,274)(518,267)
(555,261)(592,255)(611,253)(629,251)(648,249)(657,248)(667,247)(676,247)
(685,246)(690,246)(695,246)(697,246)(699,246)(702,246)(704,246)(705,246)
(706,246)(707,246)(709,246)(710,246)(711,246)(712,246)(713,246)(714,246)
(716,246)(717,246)(718,246)(720,246)(723,246)
\thinlines \path(723,246)(725,246)(727,246)(732,246)(741,247)(750,247)
(760,248)(778,250)(797,252)(816,255)(853,264)(890,275)(927,290)(964,309)
(1002,331)(1039,358)(1076,389)(1113,424)(1151,465)(1188,510)(1225,561)
(1262,617)(1300,678)(1337,746)(1374,819)(1401,877)
\thinlines \path(220,304)(220,304)(221,304)(222,304)(223,304)(225,304)
(226,304)(227,304)(228,304)(229,304)(230,304)(232,304)(233,304)(234,304)
(235,304)(236,304)(237,304)(239,304)(241,304)(243,304)(246,304)(248,304)
(250,304)(253,304)(255,304)(257,304)(262,304)(267,304)(271,304)(276,304)
(280,304)(285,304)(290,304)(294,304)(304,304)(313,304)(322,304)(332,304)
(341,304)(350,304)(360,304)(369,304)(388,305)(406,305)(425,305)(443,306)
(462,306)(481,307)(499,308)(518,309)(555,312)
\thinlines \path(555,312)(592,316)(629,321)(667,328)(704,336)(741,347)
(778,359)(816,375)(853,393)(890,414)(927,438)(964,465)(1002,497)(1039,532)
(1076,571)(1113,615)(1151,663)(1188,717)(1225,775)(1262,839)(1283,877)
\thinlines \path(220,304)(220,304)(221,304)(222,304)(223,304)(225,304)
(227,304)(229,304)(234,304)(239,303)(248,303)(257,302)(276,300)(294,297)
(332,289)(369,279)(406,269)(443,257)(481,246)(518,235)(555,224)(592,214)
(629,205)(667,198)(685,196)(704,193)(723,191)(732,191)(741,190)(746,190)
(750,190)(755,190)(757,190)(760,190)(762,190)(763,190)(764,190)(766,190)
(767,190)(768,190)(769,190)(770,190)(771,190)(773,190)(774,190)(776,190)
(778,190)(783,190)(788,190)(797,191)(806,191)
\thinlines \path(806,191)(816,192)(834,194)(853,197)(890,206)(927,218)
(964,233)(1002,253)(1039,277)(1076,305)(1113,338)(1151,376)(1188,418)
(1225,466)(1262,520)(1300,579)(1337,644)(1374,715)(1411,793)(1436,848)
\end{picture}
\vglue 2ex
\hspace*{4em}(b)$\lambda_{r}=1.25\lambda_{cr}$
\vglue 1ex
\caption{Behavior of the effective potential is shown
         at $D=2.5$
         for (a)  $\lambda_{r} < \lambda_{cr} (\lambda_{r}=0.9\lambda_{cr})$
         and (b)  $\lambda_{r} > \lambda_{cr} (\lambda_{r}=1.25\lambda_{cr}$)
         where $\lambda_{cr}=3.2/$tr 1.
         The critical curvature is given by $1/r_{cr}=\sigma_{0}/2.74$}
\label{fig:pot}
\end{figure}

To discuss the dynamical mass of the fermion
we study the minimum of this effective potential more precisely.
A necessary condition for the minimum is given by
\begin{equation}
     \left.
     \frac{\partial V(\sigma)}{\partial \sigma}
     \right|_{\sigma = m}=0\, ,
\label{def:gap}
\end{equation}
where the non-trivial solution $m$ of this equation
corresponds to the dynamical fermion mass.
Equation (\ref{def:gap}) reads
\begin{eqnarray}
     &&\frac{1}{\lambda_{r}}\sigma_{0}^{\sD-2}
     +\frac{D-1}{(4\pi)^{\sD /2}}\Gamma\left(
     1-\frac{D}{2}\right)\sigma_{0}^{\sD-2}
      \mbox{tr\boldmath$1$} \nonumber \\
     &&-\frac{r^{2-\sD}}{(4\pi)^{\sD/2}}
     \frac{\displaystyle \Gamma\left(\frac{D}{2}+i m r\right)
     \Gamma\left(\frac{D}{2}-i m r\right)}
     {\displaystyle \Gamma\left(1+i m r\right)
     \Gamma\left(1-i m r\right)}
     \Gamma\left(1-\frac{D}{2}\right)
      \mbox{tr\boldmath$1$}  \\[2mm]
     &&=0 \nonumber \, .
\label{eq:gap}
\end{eqnarray}
In Fig.2 we present the solution of the gap equation (32).

Taking the two-dimensional limit Eq.(32) reduces to
\begin{equation}
     \frac{1}{\lambda_{r}}
     -\frac{1}{2\pi}\left[\ln (r^{2}\sigma_{0}^{2})
     -\psi (1+i m r)-\psi(1-i m r)\right] = 0\, .
\end{equation}
For three-dimensions it reads \cite{EOS}
\begin{equation}
     \frac{1}{\lambda_{r}}\sigma_{0}-\frac{2}{\pi}\sigma_{0}
     +\frac{1}{\pi r}
     \frac{\displaystyle \Gamma\left(\frac{3}{2}+i m r\right)
     \Gamma\left(\frac{3}{2}-i m r\right)}
     {\displaystyle \Gamma\left(1+i m r\right)
     \Gamma\left(1-i m r\right)} =0\, ,
\label{three:gap}
\end{equation}
while it reduces to
\begin{eqnarray}
     &&\frac{1}{\lambda_{r}}
     -\frac{3}{(2\pi)^{2}}
     \left(C_{uv}-\frac{2}{3}\right)
     -\frac{2 r^{-2}}{(2\pi)^{2}\sigma_{0}^{2}} \nonumber \\
     &&+\frac{r^{-2}+m^{2}}{(2\pi)^{2}\sigma_{0}^{2}}
     \left(C_{uv}+\ln (r^{2}\sigma_{0}^{2})
     -\psi(1+i m r)-\psi(1-i m r)\right)=0\, ,
\label{gap:4d}
\end{eqnarray}
in the limit of $D\rightarrow 4$, where
\begin{equation}
     C_{uv}=\frac{2}{4-D}-\gamma+\ln 2\pi +1\, .
\end{equation}
In the weak curvature limit Eq.(\ref{three:gap}) reproduces the result
obtained in Ref.13.
For the weak curvature limit $r \rightarrow \infty$ Eq. (\ref{gap:4d}) tends to
\begin{equation}
     \frac{1}{\lambda_{r}}
     -\frac{1}{(2\pi)^{2}}\left[
     3\left(C_{uv}-\frac{2}{3}\right)
     -\left(C_{uv}-\ln\frac{m^{2}}{\sigma_{0}^{2}}\right)
     \frac{m^{2}}{\sigma_{0}^{2}}
     -\left(C_{uv}-\frac{13}{6}-\ln\frac{m^{2}}{\sigma_{0}^{2}}\right)
     \frac{1}{r^{2}\sigma_{0}^{2}}\right]
     =0\, .
\label{gap:4dweak}
\end{equation}
We find that
there is a simple correspondence between this result (\ref{gap:4dweak})
and the result given
in Ref.14 if we make a replacement
\begin{equation}
     C_{uv}+\frac{1}{6}\frac{1}{(m^{2}-3\sigma_{0}^{2})r^{2}}
     \leftrightarrow \ln \frac{\Lambda^{2}}{\sigma_{0}^{2}}\, ,
\end{equation}
where $\Lambda$ is a cut-off of the divergent integral appearing in Ref.14.
Note here that the direct comparison of our result
with the result in Ref.14 is possible only after renormalizing
the coupling constant $\lambda$ in Ref.14 under the renormalization
condition (\ref{cond:ren}).
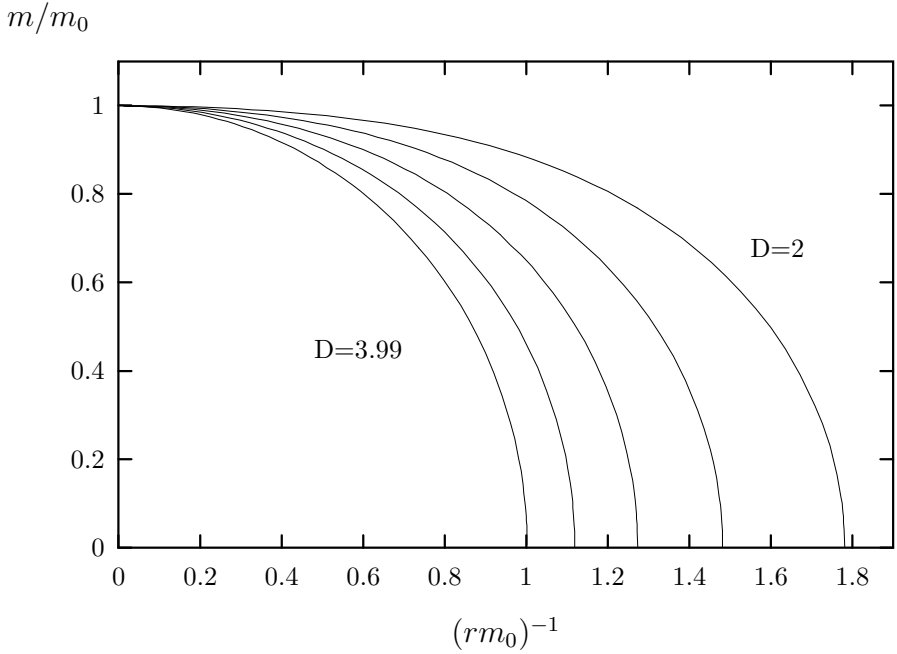
\begin{figure}[H]
% GNUPLOT: LaTeX picture using EEPIC macros
\setlength{\unitlength}{0.240900pt}
\begin{picture}(1500,900)(0,0)
\tenrm
\thicklines \path(220,113)(240,113)
\thicklines \path(1436,113)(1416,113)
\put(198,113){\makebox(0,0)[r]{0}}
\thicklines \path(220,252)(240,252)
\thicklines \path(1436,252)(1416,252)
\put(198,252){\makebox(0,0)[r]{0.2}}
\thicklines \path(220,391)(240,391)
\thicklines \path(1436,391)(1416,391)
\put(198,391){\makebox(0,0)[r]{0.4}}
\thicklines \path(220,530)(240,530)
\thicklines \path(1436,530)(1416,530)
\put(198,530){\makebox(0,0)[r]{0.6}}
\thicklines \path(220,669)(240,669)
\thicklines \path(1436,669)(1416,669)
\put(198,669){\makebox(0,0)[r]{0.8}}
\thicklines \path(220,808)(240,808)
\thicklines \path(1436,808)(1416,808)
\put(198,808){\makebox(0,0)[r]{1}}
\thicklines \path(220,113)(220,133)
\thicklines \path(220,877)(220,857)
\put(220,68){\makebox(0,0){0}}
\thicklines \path(348,113)(348,133)
\thicklines \path(348,877)(348,857)
\put(348,68){\makebox(0,0){0.2}}
\thicklines \path(476,113)(476,133)
\thicklines \path(476,877)(476,857)
\put(476,68){\makebox(0,0){0.4}}
\thicklines \path(604,113)(604,133)
\thicklines \path(604,877)(604,857)
\put(604,68){\makebox(0,0){0.6}}
\thicklines \path(732,113)(732,133)
\thicklines \path(732,877)(732,857)
\put(732,68){\makebox(0,0){0.8}}
\thicklines \path(860,113)(860,133)
\thicklines \path(860,877)(860,857)
\put(860,68){\makebox(0,0){1}}
\thicklines \path(988,113)(988,133)
\thicklines \path(988,877)(988,857)
\put(988,68){\makebox(0,0){1.2}}
\thicklines \path(1116,113)(1116,133)
\thicklines \path(1116,877)(1116,857)
\put(1116,68){\makebox(0,0){1.4}}
\thicklines \path(1244,113)(1244,133)
\thicklines \path(1244,877)(1244,857)
\put(1244,68){\makebox(0,0){1.6}}
\thicklines \path(1372,113)(1372,133)
\thicklines \path(1372,877)(1372,857)
\put(1372,68){\makebox(0,0){1.8}}
\thicklines \path(220,113)(1436,113)(1436,877)(220,877)(220,113)
\put(45,945){\makebox(0,0)[l]{\shortstack{$m/m_{0}$}}}
\put(828,-22){\makebox(0,0){$(r m_{0})^{-1}$}}
\put(1212,585){\makebox(0,0)[l]{D=2}}
\put(527,426){\makebox(0,0)[l]{D=3.99}}
\thinlines \path(220,808)(220,808)(221,808)(222,808)(222,808)(223,808)
(224,808)(226,808)(227,808)(229,808)(232,808)(235,808)(238,808)(243,807)
(249,807)(255,807)(267,807)(278,807)(290,807)(313,806)(336,806)(360,805)
(383,804)(406,803)(429,801)(453,800)(476,798)(499,796)(522,794)(546,792)
(569,789)(592,786)(616,783)(639,780)(662,776)(685,772)(709,767)(732,762)
(755,757)(778,751)(802,745)(825,738)(848,731)(871,723)(895,714)(918,705)
(941,695)(964,684)(988,673)(1011,660)(1034,647)
\thinlines \path(1034,647)(1057,632)(1081,616)(1104,600)(1127,581)(1151,561)
(1174,539)(1197,515)(1220,489)(1244,459)(1267,425)(1290,385)(1313,337)
(1325,308)(1337,273)(1342,252)(1348,227)(1354,193)(1356,183)(1357,170)
(1358,162)(1358,153)(1359,142)(1360,115)
\thinlines \path(220,808)(220,808)(221,808)(221,808)(222,808)(222,808)
(224,808)(225,808)(226,808)(227,808)(230,808)(232,808)(235,807)(239,807)
(244,807)(249,807)(259,807)(268,807)(278,807)(297,806)(317,805)(336,804)
(356,803)(375,801)(394,799)(414,797)(433,795)(452,793)(472,790)(491,787)
(510,784)(530,780)(549,777)(568,773)(588,768)(607,764)(626,758)(646,753)
(665,747)(684,741)(704,734)(723,726)(743,719)(762,710)(781,701)(801,692)
(820,681)(839,670)(859,659)(878,646)(897,632)
\thinlines \path(897,632)(917,618)(936,602)(955,585)(975,567)(994,547)
(1013,526)(1033,502)(1052,476)(1071,447)(1091,414)(1110,376)(1129,329)
(1139,301)(1149,267)(1154,247)(1159,222)(1163,191)(1165,180)(1166,168)
(1167,152)(1168,140)(1168,114)
\thinlines \path(220,808)(220,808)(221,808)(221,808)(222,808)(222,808)
(223,808)(224,808)(225,808)(226,808)(228,808)(233,807)(237,807)(241,807)
(245,807)(253,807)(262,807)(270,806)(287,806)(303,805)(320,803)(336,802)
(353,800)(370,798)(386,796)(403,793)(420,790)(436,787)(453,784)(469,780)
(486,776)(503,772)(519,767)(536,762)(553,757)(569,751)(586,745)(603,739)
(619,732)(636,725)(652,717)(669,708)(686,700)(702,690)(719,680)(736,670)
(752,658)(769,646)(785,633)(802,620)(819,605)
\thinlines \path(819,605)(835,589)(852,573)(869,555)(885,535)(902,514)
(918,491)(935,465)(952,437)(968,405)(985,367)(1002,322)(1010,295)(1018,262)
(1022,242)(1027,219)(1029,205)(1031,188)(1032,178)(1033,166)(1033,159)
(1034,150)(1034,139)(1035,114)
\thinlines \path(220,808)(220,808)(221,808)(221,808)(221,808)(222,808)
(223,808)(224,808)(226,808)(227,808)(229,807)(231,807)(235,807)(238,807)
(242,807)(249,807)(257,807)(264,806)(279,805)(293,804)(308,803)(322,801)
(337,799)(352,797)(366,794)(381,791)(395,788)(410,785)(425,781)(439,777)
(454,772)(469,768)(483,763)(498,757)(512,752)(527,745)(542,739)(556,732)
(571,725)(585,717)(600,709)(615,700)(629,691)(644,681)(659,671)(673,660)
(688,648)(702,636)(717,623)(732,609)(746,594)
\thinlines \path(746,594)(761,578)(775,561)(790,543)(805,524)(819,503)
(834,480)(849,455)(863,427)(878,396)(892,359)(907,315)(914,289)(922,257)
(925,238)(929,215)(931,202)(933,185)(933,176)(934,164)(935,149)(936,139)
(936,114)
\thinlines \path(220,808)(220,808)(220,808)(221,808)(221,808)(222,808)
(223,808)(223,808)(225,808)(227,808)(228,807)(230,807)(233,807)(236,807)
(240,807)(246,807)(253,807)(259,806)(272,805)(286,804)(299,802)(312,800)
(325,798)(338,796)(351,793)(364,790)(377,786)(390,783)(403,779)(416,774)
(429,770)(443,764)(456,759)(469,753)(482,747)(495,741)(508,734)(521,726)
(534,719)(547,710)(560,702)(573,693)(587,683)(600,673)(613,662)(626,651)
(639,639)(652,627)(665,613)(678,599)(691,584)
\thinlines \path(691,584)(704,569)(717,552)(730,534)(744,514)(757,494)
(770,471)(783,446)(796,419)(809,388)(822,352)(835,310)(842,284)(848,253)
(852,234)(855,212)(856,199)(858,183)(859,174)(860,163)(860,156)(861,148)
(861,138)(861,117)
\end{picture}
\caption{Behavior of the Dynamical fermion mass as a function of the
         radius $r$ at $D=2.0,2.5,3.0,3.5,3.99$ where $m_{0}$
         is the dynamical fermion mass in flat space-time.}
\label{fig:mass}
\end{figure}

As is seen in Fig.1 and Fig.2 the phase transition is of the second order.
Accordingly by setting $m=0$ in the gap equation (32)
we may derive the equation which determines
the critical radius $r_{cr}$,
\begin{equation}
     \frac{1}{\lambda_{r}}\sigma_{0}^{\sD-2}
     +\frac{D-1}{(4\pi)^{\sD /2}}\Gamma\left(
     1-\frac{D}{2}\right)\sigma_{0}^{\sD-2} \mbox{tr\boldmath$1$}
     -\frac{r_{cr}^{2-\sD}}{(4\pi)^{\sD /2}}
     \Gamma^{2}\left(\frac{D}{2}\right)
     \Gamma\left(1-\frac{D}{2}\right)
      \mbox{tr\boldmath$1$} =0\, .
\label{eq:phase}
\end{equation}
Hence the critical radius is given by
\begin{equation}
     r_{cr}=\frac{1}{\sigma_{0}}\left[
     \frac{(4\pi)^{\sD/2}}
     {\displaystyle
      \Gamma^2\left(\frac{D}{2}\right)\Gamma\left(1-\frac{D}{2}\right)
     \mbox{tr\boldmath$1$}}
     \frac{1}{\lambda_{r}}
     +\frac{D-1}{\displaystyle\Gamma^{2}\left(\frac{D}{2}\right)}
     \right]^{1/(2-D)}\, .
\label{r:cr}
\end{equation}
For some special values of $D$ Eq.(\ref{r:cr}) simplifies to:
\begin{equation}
\left\{
\begin{array}{ll}
     {\displaystyle r_{cr}=\frac{1}{\sigma_{0}}
     \exp\left(\frac{\pi}{\lambda_{r}}-1-\gamma\right)}\,  &; D=2\, ,\\[4mm]
     {\displaystyle r_{cr}=\frac{1}{\sigma_{0}}\left(\frac{8}{\pi}
     -4\frac{1}{\lambda_{r}}\right)^{-1}}\,  &; D=3\, ,\\[4mm]
     {\displaystyle r_{cr}=\frac{\sqrt{3}}{3}\frac{1}{\sigma_{0}}}
     \,  &; D=4\, .
\end{array}
\right.
\end{equation}
In Fig.3 we show the critical radius $r_{cr}$ as a function of
the coupling constant $\lambda_{r}$.
\newpage
\begin{figure}[H]
% GNUPLOT: LaTeX picture using EEPIC macros
\setlength{\unitlength}{0.240900pt}
\begin{picture}(1500,900)(0,0)
\tenrm
\thicklines \path(220,113)(240,113)
\thicklines \path(1436,113)(1416,113)
\put(198,113){\makebox(0,0)[r]{0}}
\thicklines \path(220,249)(240,249)
\thicklines \path(1436,249)(1416,249)
\put(198,249){\makebox(0,0)[r]{0.5}}
\thicklines \path(220,386)(240,386)
\thicklines \path(1436,386)(1416,386)
\put(198,386){\makebox(0,0)[r]{1}}
\thicklines \path(220,522)(240,522)
\thicklines \path(1436,522)(1416,522)
\put(198,522){\makebox(0,0)[r]{1.5}}
\thicklines \path(220,659)(240,659)
\thicklines \path(1436,659)(1416,659)
\put(198,659){\makebox(0,0)[r]{2}}
\thicklines \path(220,795)(240,795)
\thicklines \path(1436,795)(1416,795)
\put(198,795){\makebox(0,0)[r]{2.5}}
\thicklines \path(220,113)(220,133)
\thicklines \path(220,877)(220,857)
\put(220,68){\makebox(0,0){0}}
\thicklines \path(463,113)(463,133)
\thicklines \path(463,877)(463,857)
\put(463,68){\makebox(0,0){1}}
\thicklines \path(706,113)(706,133)
\thicklines \path(706,877)(706,857)
\put(706,68){\makebox(0,0){2}}
\thicklines \path(950,113)(950,133)
\thicklines \path(950,877)(950,857)
\put(950,68){\makebox(0,0){3}}
\thicklines \path(1193,113)(1193,133)
\thicklines \path(1193,877)(1193,857)
\put(1193,68){\makebox(0,0){4}}
\thicklines \path(1436,113)(1436,133)
\thicklines \path(1436,877)(1436,857)
\put(1436,68){\makebox(0,0){5}}
\thicklines \path(220,113)(1436,113)(1436,877)(220,877)(220,113)
\put(45,945){\makebox(0,0)[l]{\shortstack{$(r_{cr}\sigma_{0})^{-1}$}}}
\put(828,-22){\makebox(0,0){$\lambda_{r}$}}
\put(658,522){\makebox(0,0)[l]{D=2}}
\put(950,495){\makebox(0,0)[l]{D=2.5}}
\put(1071,318){\makebox(0,0)[l]{D=3}}
\thinlines \path(220,113)(220,113)(221,113)(222,113)(222,113)(223,113)
(224,113)(225,113)(225,113)(226,113)(227,113)(228,113)(229,113)(229,113)
(230,113)(231,113)(232,113)(232,113)(233,113)(234,113)(235,113)(236,113)
(236,113)(237,113)(238,113)(239,113)(239,113)(240,113)(241,113)(242,113)
(242,113)(243,113)(244,113)(245,113)(246,113)(246,113)(247,113)(248,113)
(249,113)(249,113)(250,113)(251,113)(252,113)(253,113)(253,113)(254,113)
(255,113)(256,113)(256,113)(257,113)(258,113)
\thinlines \path(258,113)(259,113)(260,113)(260,113)(261,113)(262,113)
(263,113)(263,113)(264,113)(265,113)(266,113)(267,113)(267,113)(268,113)
(270,113)(270,113)(271,113)(272,113)(273,113)(274,113)(276,113)(277,113)
(279,113)(280,113)(282,113)(284,113)(285,113)(287,113)(288,113)(290,113)
(291,113)(294,113)(298,113)(301,113)(304,113)(307,113)(310,113)(313,113)
(316,113)(319,114)(325,114)(332,114)(338,115)(344,116)(356,118)(369,121)
(381,125)(394,129)(419,141)(443,156)(468,174)
\thinlines \path(468,174)(493,193)(518,215)(543,237)(567,260)(592,283)
(617,306)(642,329)(667,352)(692,374)(716,396)(741,418)(766,439)(791,459)
(816,479)(840,499)(865,517)(890,535)(915,553)(940,570)(964,586)(989,602)
(1014,618)(1039,633)(1064,647)(1089,661)(1113,675)(1138,688)(1163,701)
(1188,713)(1213,725)(1237,736)(1262,748)(1287,759)(1312,769)(1337,779)
(1362,789)(1386,799)(1411,809)(1436,818)
\thinlines \path(549,113)(549,113)(550,113)(550,113)(551,113)(551,113)
(552,113)(554,113)(556,113)(558,114)(563,114)(567,115)(576,118)(585,122)
(603,131)(621,143)(640,155)(658,169)(676,183)(694,198)(712,213)(730,228)
(748,242)(766,257)(784,271)(802,285)(821,299)(839,312)(857,325)(875,338)
(893,351)(911,363)(929,374)(947,386)(965,397)(983,407)(1002,418)(1020,428)
(1038,438)(1056,447)(1074,457)(1092,466)(1110,474)(1128,483)(1146,491)
(1164,499)(1183,507)(1201,515)(1219,522)(1237,529)(1255,536)
\thinlines \path(1255,536)(1273,543)(1291,550)(1309,556)(1327,562)(1345,568)
(1364,574)(1382,580)(1400,586)(1418,592)(1436,597)
\thinlines \path(760,113)(760,113)(774,130)(788,147)(802,162)(815,177)
(829,192)(843,205)(857,218)(871,231)(884,243)(898,254)(912,265)(926,276)
(940,286)(953,296)(967,305)(981,314)(995,323)(1008,332)(1022,340)(1036,348)
(1050,355)(1064,363)(1077,370)(1091,377)(1105,384)(1119,390)(1133,396)
(1146,403)(1160,409)(1174,414)(1188,420)(1202,425)(1215,431)(1229,436)
(1243,441)(1257,446)(1271,450)(1284,455)(1298,460)(1312,464)(1326,468)
(1339,473)(1353,477)(1367,481)(1381,484)(1395,488)(1408,492)(1422,496)
(1436,499)
\end{picture}
\vglue 2ex
\hspace*{4em}(a)Critical radius $r_{cr}$ at $D=2.0,2.5,3.0$
\vglue 6ex
% GNUPLOT: LaTeX picture using EEPIC macros
\setlength{\unitlength}{0.240900pt}
\begin{picture}(1500,900)(0,0)
\tenrm
\thicklines \path(220,113)(240,113)
\thicklines \path(1436,113)(1416,113)
\put(198,113){\makebox(0,0)[r]{0}}
\thicklines \path(220,198)(240,198)
\thicklines \path(1436,198)(1416,198)
\put(198,198){\makebox(0,0)[r]{0.2}}
\thicklines \path(220,283)(240,283)
\thicklines \path(1436,283)(1416,283)
\put(198,283){\makebox(0,0)[r]{0.4}}
\thicklines \path(220,368)(240,368)
\thicklines \path(1436,368)(1416,368)
\put(198,368){\makebox(0,0)[r]{0.6}}
\thicklines \path(220,453)(240,453)
\thicklines \path(1436,453)(1416,453)
\put(198,453){\makebox(0,0)[r]{0.8}}
\thicklines \path(220,537)(240,537)
\thicklines \path(1436,537)(1416,537)
\put(198,537){\makebox(0,0)[r]{1}}
\thicklines \path(220,622)(240,622)
\thicklines \path(1436,622)(1416,622)
\put(198,622){\makebox(0,0)[r]{1.2}}
\thicklines \path(220,707)(240,707)
\thicklines \path(1436,707)(1416,707)
\put(198,707){\makebox(0,0)[r]{1.4}}
\thicklines \path(220,792)(240,792)
\thicklines \path(1436,792)(1416,792)
\put(198,792){\makebox(0,0)[r]{1.6}}
\thicklines \path(220,877)(240,877)
\thicklines \path(1436,877)(1416,877)
\put(198,877){\makebox(0,0)[r]{1.8}}
\thicklines \path(220,113)(220,133)
\thicklines \path(220,877)(220,857)
\put(220,68){\makebox(0,0){0}}
\thicklines \path(463,113)(463,133)
\thicklines \path(463,877)(463,857)
\put(463,68){\makebox(0,0){1}}
\thicklines \path(706,113)(706,133)
\thicklines \path(706,877)(706,857)
\put(706,68){\makebox(0,0){2}}
\thicklines \path(950,113)(950,133)
\thicklines \path(950,877)(950,857)
\put(950,68){\makebox(0,0){3}}
\thicklines \path(1193,113)(1193,133)
\thicklines \path(1193,877)(1193,857)
\put(1193,68){\makebox(0,0){4}}
\thicklines \path(1436,113)(1436,133)
\thicklines \path(1436,877)(1436,857)
\put(1436,68){\makebox(0,0){5}}
\thicklines \path(220,113)(1436,113)(1436,877)(220,877)(220,113)
\put(45,945){\makebox(0,0)[l]{\shortstack{$(r_{cr}\sigma_{0})^{-1}$}}}
\put(828,-22){\makebox(0,0){$\lambda_{r}$}}
\put(463,771){\makebox(0,0)[l]{D=3.99}}
\put(706,495){\makebox(0,0)[l]{D=3.5}}
\put(1071,431){\makebox(0,0)[l]{D=3}}
\thinlines \path(760,113)(760,113)(774,140)(788,165)(802,190)(815,213)
(829,235)(843,257)(857,277)(871,296)(884,315)(898,333)(912,350)(926,366)
(940,382)(953,398)(967,412)(981,426)(995,440)(1008,453)(1022,466)(1036,478)
(1050,490)(1064,502)(1077,513)(1091,524)(1105,534)(1119,544)(1133,554)
(1146,564)(1160,573)(1174,582)(1188,590)(1202,599)(1215,607)(1229,615)
(1243,623)(1257,631)(1271,638)(1284,645)(1298,652)(1312,659)(1326,666)
(1339,672)(1353,679)(1367,685)(1381,691)(1395,697)(1408,702)(1422,708)
(1436,714)
\thinlines \path(722,113)(722,113)(722,121)(723,126)(723,130)(724,134)
(725,140)(725,146)(729,165)(736,194)(751,239)(766,276)(780,307)(795,334)
(809,358)(824,380)(838,401)(853,419)(868,437)(882,453)(897,468)(911,482)
(926,496)(940,508)(955,520)(970,531)(984,542)(999,552)(1013,562)(1028,571)
(1042,580)(1057,588)(1072,596)(1086,604)(1101,612)(1115,619)(1130,626)
(1144,632)(1159,639)(1174,645)(1188,651)(1203,656)(1217,662)(1232,667)
(1247,672)(1261,677)(1276,682)(1290,687)(1305,691)(1319,696)(1334,700)
\thinlines \path(1334,700)(1349,704)(1363,708)(1378,712)(1392,716)(1407,720)
(1421,723)(1436,727)
\thinlines \path(236,113)(236,113)(237,270)(237,330)(239,408)(240,437)
(240,461)(242,501)(244,532)(245,558)(248,599)(251,629)(254,652)(257,671)
(260,687)(266,711)(273,729)(279,743)(285,753)(297,770)(309,781)(322,790)
(334,797)(346,802)(358,807)(383,813)(407,818)(432,822)(456,825)(481,828)
(505,830)(530,831)(554,833)(579,834)(603,835)(628,836)(652,837)(677,838)
(701,838)(726,839)(750,839)(775,840)(799,840)(824,841)(848,841)(873,841)
(897,842)(922,842)(946,842)(971,843)(995,843)
\thinlines \path(995,843)(1020,843)(1044,843)(1069,844)(1093,844)(1118,844)
(1142,844)(1167,844)(1191,844)(1216,845)(1240,845)(1265,845)(1289,845)
(1314,845)(1338,845)(1363,845)(1387,845)(1412,846)(1436,846)
\end{picture}
\vglue 2ex
\hspace*{4em}(b)Critical radius $r_{cr}$ at $D=3.0,3.5,3.99$
\vglue 1ex
\caption{Critical radius $r_{cr}$ as a function of
four-fermion coupling $\lambda_{r}$}
\label{fig:phase2}
\end{figure}

\newpage

We found that the phase structure of the four-fermion interaction theory in
de Sitter space is analyzable in the sense of the 1/N expansion and
discovered an existance of the critical curvature at which the symmetry
is restored.
Although our model is too primitive to be adopted to the symmetry
restoration of the unified theories in early universe,
we hope that our analysis will help building a more realistic composite
Higgs model in early stage of the universe.

We would like to thank K.~Fukazawa, T.~Inami, K.~Ishikawa, S.~D.~Odintsov,
H.~Sato and H.~Suzuki
for discussions and useful information.

\end{document}